\documentclass[prl,twocolumn,nofootinbib]{revtex4-2}
\usepackage{amsmath,amssymb,amsfonts}
\usepackage{xcolor}

\newcommand{\beq}{\begin{eqnarray}}
\newcommand{\eeq}{\end{eqnarray}}
\newcommand{\beqn}{\begin{eqnarray}}
\newcommand{\eeqn}{\end{eqnarray}}
\newcommand{\be}{\begin{equation}}
\newcommand{\ee}{\end{equation}}
\newcommand{\bea}{\begin{eqnarray}}
\newcommand{\eea}{\end{eqnarray}}

\newcommand{\ZZ}{\mathbb{Z}}

\begin{document}
\newcommand{\thistitle}{The Vacuum Energy Density and Gravitational Entropy}
\newcommand{\uiuc}{Illinois Center for Advanced Studies of the Universe \& Department of Physics, University of Illinois, 1110 West Green St., Urbana IL 61801, U.S.A.}
\newcommand{\perimeter}{Perimeter Institute for Theoretical Physics, 31 Caroline St. N., Waterloo ON, Canada}
\newcommand{\vt}{Department  of Physics, Virginia Tech, Blacksburg, VA 24061, U.S.A.}
\newcommand{\poland}{University of Wroc\l{}aw, Faculty of Physics and Astronomy, pl.\ M.\ Borna 9, 50-204 Wroc\l{}aw, Poland\\
National Centre for Nuclear Research, Pasteura 7, 02-093 Warsaw, Poland}

\title{\thistitle}
\author{Laurent Freidel}
\affiliation{\perimeter}
\author{Jerzy Kowalski-Glikman}
\affiliation{\poland}
\author{Robert G. Leigh}
\affiliation{\uiuc}
\author{Djordje Minic}
\affiliation{\vt}

\begin{abstract}
The failure to calculate the vacuum energy is a central problem in theoretical physics. Presumably the problem arises from the insistent use of effective field theory reasoning in a context that is well beyond its intended scope. If one follows this path, one is led inevitably to statistical or anthropic reasoning for observations. It appears that a more palatable resolution of the vacuum energy problem requires some form of UV/IR feedback. In this paper we take the point of view that such feedback can be thought of as arising by  defining a notion of quantum space-time. 
We reformulate the regularized computation of vacuum energy in such a way that it can be interpreted in terms of a sum over elementary phase space volumes, that we identify with a ground state degeneracy. This observation yields a precise notion of UV/IR feedback, while leaving a scale unfixed.  Here we argue that holography can be thought to provide a key piece of information: 
%
we show that equating this microscopic ground state degeneracy with macroscopic gravitational entropy  yields a prediction for the vacuum energy that can easily be consistent with observations. Essentially, the smallness of the vacuum energy is tied to the large size of the Universe. We discuss how within this scenario notions of effective field theory can go so wrong.
\end{abstract}
\maketitle

The success of the standard models of particle physics and of cosmology are unquestionable. And yet, vexing problems remain, often associated with apparent hierarchies of scales. In the absence of gravity in our description, effective field theory (EFT) methods are expected to apply and in fact are foundational in the way that we think of quantum field theoretic physics \cite{Baumgart:2022yty}. In recent years however, it has become increasingly clear that the nature of the vacuum in gauge theories and gravity is much richer than previously thought \cite{Strominger:2017zoo}, \cite{Donnelly:2016auv} (see also, \cite{Freidel:2021wpl}, \cite{Berglund:2022qcc}). In addition, it is without a doubt that holography and related quantum information theoretic ideas are important ingredients in our understanding of gravity and are expected to play a central role in a quantum theory \cite{Harlow:2022qsq}. It is natural to ask if each of these concepts are important for the central gravitational hierarchy problem, that of the vacuum energy density, and if they complicate the application of EFT methods \cite{Cohen:1998zx}, \cite{Donoghue:2020hoh}.

There are many extant reviews on the vacuum energy density problem 
\cite{Weinberg:1988cp, Weinberg:2000yb, Witten:2000zk, Padmanabhan:2002ji, Polchinski:2006gy}. 
The basic prediction (of effective field theory) is that this scales as the fourth power of a momentum space cutoff. This prediction differs exponentially from the observed value unless that cutoff is significantly smaller than expected. It is also notable that perturbative string theory seemingly ameliorates the problem through its ultraviolet finiteness (certainly a non-EFT property) 
and yet offers no solution as an arbitrary cutoff is simply replaced by the string mass scale, 
as pointed out by Polchinski long ago \cite{Polchinski:1998rq, Polchinski:1985zf}. 
Much effort has gone into explaining away the problem by anthropic or statistical reasoning but to most these ideas 
\cite{Weinberg:2000yb, Polchinski:2006gy}, if they could be convincingly implemented, would be disappointing. Clearly, a resolution of the problem within traditional EFT is impossible. As mentioned briefly by Polchinski in his review 
\cite{Polchinski:2006gy} and by others, the vacuum energy density problem seems to point to some feedback mechanism between the ultra-violet (UV) and infra-red (IR). An implementation of this mechanism, as far as we are aware, has never been offered. 

In this paper, we will introduce such a mechanism. In a few words, using just prime principles of quantum theory,  we will introduce a way to enumerate the vacua of a microscopic theory. Essentially this constitutes a novel regulator for observables (such as the vacuum energy density) that directly address vacua and it can in particular be employed to calculate a regulated value of the vacuum energy density. This calculation involves an undetermined length scale; this scale can be fixed by equating the microscopic count of vacua with the macroscopic value of gravitational entropy. Remarkably, using this information for the regulated vacuum energy density yields a prediction that is consistent with observation.
Heuristically one can say that the holographic bound on entropy precludes the size of momentum space (and thus effectively its cutoff) being as large as EFT assumes. These ideas originate in the realization that the trace involved in the computation of the vacuum energy can be understood in terms of {\it phase space geometry}. The vacua referred to above correspond to elementary quantum areas in phase space. A prime principle of quantum theory is the basis independence of a trace; in the present context, a judicious choice of basis (called a modular basis in \cite{aharonov, Freidel:2016pls})  makes direct contact with minimal phase space areas. Remarkably, this quantum area  constraint translates to a link between UV and IR scales.

The standard calculation of the vacuum energy density proceeds by computing the circle amplitude. This can be thought of directly in Feynman diagram terms but since the diagram has no external legs and involves no interactions at one loop, is equally well expressible using the world-line formalism, yielding\footnote{See for example  Section 7.3  of \cite{Polchinski:1998rq}, pages 216-222. }
\beq
Z_{S^1}=\int_0^\infty \frac{d\tau}{2\tau} \mathrm{Tr}\, e^{i\hat{\cal H}\tau} 
\eeq
where in this expression $\hat{\cal H}$ is the Hamiltonian constraint and $\tau$ the intrinsic length of the world-line. As a trace, we expect the value of $Z_{S^1}$ to be basis independent. A choice of basis is equivalent to a choice of polarization of phase space. 
In particle theory, it is natural to choose the momentum basis, and we would write\footnote{By $\delta^{(4)}(0)$, we mean the limit $\lim_{q\to 0} \delta^{(4)}(q)$ of a delta function in momentum space.}
\beqn
Z_{S^1}&=&\int_0^\infty \frac{d\tau}{2\tau}\int\frac{d^4p}{(2\pi)^4} \langle p_\mu| e^{i\hat{\cal H}(\hat p)\tau}| p_\mu\rangle\\
&=&\delta^{(4)}(0)\int_0^\infty \frac{d\tau}{2\tau}\int \frac{d^4p}{(2\pi)^4} e^{i{\cal H}( p)\tau}.
\eeqn
We see that this choice of basis yields a singular result. The standard way to deal with this is well-known: one interprets the momentum space $\delta$-function as the space-time volume,
\beq
Z_{S^1}
=\int_0^\infty \frac{d\tau}{2\tau}\int \frac{d^4q d^4p}{(2\pi\hbar)^4}\, e^{i{\cal H}( p)\tau}.\label{vacenergy}
\eeq
We note that this casts the trace as an integration over the classical phase space. The usual treatment is to take the space-time volume $V_q=\int d^4q$ as fixed but finite and treat the momentum and $\tau$ integrals separately (which require further regularization). It should be emphasized that this step explicitly uses a standard but singular polarization of phase space: the Schr\"odinger polarization in which the fundamental phase space cell is collapsed onto a line \cite{Freidel:2016pls}.

Integrating over the modular parameter $\tau$ reduces the trace to a sum over {\it on-shell } particle states and gives the familiar result
\beq
\int_0^\infty \frac{d\tau}{2\tau}V_q\int\frac{d^4p}{(2\pi\hbar)^4} e^{i{\cal H}(p)\tau}\to V_q\int\frac{d^3p}{(2\pi\hbar)^3}\frac12\omega_{\vec p}
\eeq
where we have taken ${\cal H}(p)=\frac12(p_\mu^2+m^2)$ and $ \hbar \omega_{\vec{p}}=\sqrt{\vec p^2+m^2}$. 
Recall that the one-loop vacuum amplitude is related to the circle amplitude by exponentiation, $
Z_{vac}=\exp(Z_{S^1})=\langle e^{-i\hat H T/\hbar}\rangle,
$
yielding     
\beq\label{relZrho}
Z_{S^1}=\rho V_q/\hbar,
\eeq
where $\rho$ is the vacuum energy density 
\cite{Polchinski:1998rq}. 

As usual, this integral is UV divergent, and implementing a simple cutoff yields a scaling as the cutoff to the fourth power \cite{Donoghue:2020hoh}. We emphasize here that in this computation, there is an implicit assumption of a unique spatially translation-invariant ground state, with the momentum $\vec p$ thought of as a label on the physical particle states that propagate in a fixed space. This is a possible interpretation once the $\tau$-integration has been {done}, reducing the trace to a sum over on-shell particle states. If we had several species of particles, we would continue with this interpretation, simply summing over the species. 

In perturbative string theory \cite{Polchinski:1998rq, Polchinski:1985zf}, the integral over $p$ comes from the zero modes of the string in flat space-time; in the limit of a long thin string, the torus amplitude has a traditional interpretation as an (infinite) sum over particle states, again with $p_\mu$ interpreted as the momentum of each off-shell particle state. In string theory, the $\tau$ integral is also replaced by an integral over the moduli of  the torus, which can be thought of as a complexification of $\tau$.  

Having reviewed the standard approach to the vacuum energy problem, let us return to the formula \eqref{vacenergy}. In the analogous calculation in string theory, one leaves the integral over moduli to the end. It is convenient to  do the same here in the field theory computation, focusing on the integral
\beq
Z(\tau)=\int\frac{d^4qd^4p}{(2\pi \hbar)^4} e^{-p_\mu^2\tau/2}=\, \mathrm{Tr}\, e^{-\hat p_{\mu}^2\tau/2},
\eeq
where we have Wick rotated  to Euclidean signature and similarly redefined $\tau$, and 
\beq
Z_{S^1}=\int_0^\infty \frac{d\tau}{2\tau}e^{-m^2\tau} Z(\tau).
\eeq
By doing so, we are free to regard this trace directly, without an assumed interpretation in terms of particle states -- it is simply a trace in (4 copies of) the Heisenberg group Hilbert space. 
In what follows, we will make use of the expectation that the volume in phase space is related to a count of the number of degrees of freedom (that is, an entropy), and indeed we will suggest that the vacuum energy is bounded by this count. 

To proceed, note that we may write
\beqn
Z(\tau)&=&\prod_{j=1}^4\Big[ \frac{1}{2\pi\hbar}\int_{-\infty}^\infty dq_j\int_{-\infty}^\infty dp_j\, e^{-p_j^2\tau/2}\Big]\\
&=&\Big[ \frac{\lambda\varepsilon}{2\pi\hbar}\sum_{k,\tilde k\in\ZZ}\int_0^1 dx\int_{0}^1d\tilde x\, e^{-(\tilde x+k)^2\varepsilon^2\tau/2}\Big]^4
\label{modpolform}
\eeqn
Here we have split the integral over phase space into a sum over integrals in a finite cell, via $p\to \varepsilon\tilde x$, $q\to\lambda x$. The dimensionful scales $\lambda,\varepsilon$ are arbitrary here. While the manipulations done above are trivial rearrangements of the integral, the result on the second line has an interpretation of having done the trace in another basis, a so-called {\it modular polarization} \cite{Freidel:2016pls, Freidel:2015uug, Freidel:2015pka}, which is unitarily equivalent (via Zak transform) to the momentum basis. One refers to $(x,\tilde x)\in[0,1]^2$ as a {\it modular cell} \cite{Freidel:2016pls}. The sums over $k,\tilde k$ can then be interpreted as counting such modular cells. In fact it is possible to carry out the quantization procedure directly in the modular polarization \cite{Freidel:2016pls}, and one finds that there is one state per cell, characterized by a function possessing one zero per modular cell. 
Rewriting $Z(\tau)$ in the form \eqref{modpolform} suggests that the modular cell should be interpreted as the fundamental minimal area cell in phase space. That is, we identify
\beq\label{quarea}
\lambda\varepsilon = 2\pi\hbar,
\eeq
which we refer to as the quantum area constraint. 
Whereas the area of the modular cell is fixed by this quantum constraint, its shape is not.\footnote{More generally one could consider an arbitrary tiling of phase space, with local values of $\lambda,\varepsilon$. In this paper, we make a simplifying choice of homogeneity.} One might expect the shape to be determined contextually.

Now of course $Z(\tau)$ is divergent and so we must introduce a regulator to deal with it further. Here, we will simply restrict the sums over $k,\tilde k$ to a finite range, and write 
\beqn\label{fullZtaumr}
Z(\tau)_{m.r.}:=\Big[
\sum_{k=0}^{N_{q}-1}\sum_{\tilde k=0}^{N_{p}-1}\int_0^1 dxd\tilde x\, e^{-(\tilde x+k)^2\varepsilon^2\tau/2}\Big]^4
\eeqn
where $N_p$ and $N_q$ are finite integers. We refer to this as a {\it modular regularization}, but we note that given $\lambda,\varepsilon$, then $N_q$ and $N_p$ determine the total spatial and momentum size of the regulated phase space
\beq\label{bigscales}
L:= N_q\lambda,\qquad M:= N_p\varepsilon.
\eeq
From this point of view, we haven't done anything non-standard and have merely cutoff the size of both space and momentum. However, given the interpretation of the modular cell as a minimal area cell in phase space, and assigning one quantum degree of freedom to each cell, we can associate $N_q,N_p$ with the total number of degrees of freedom 
\beq\label{grdstdegen}
N= (N_qN_p)^4
\eeq
in four space-time dimensions. This allows us to rewrite $(\lambda,\varepsilon,N_q,N_p)$ in terms of $(L,N,\lambda,\hbar)$. 
We note in particular that combining \eqref{grdstdegen} with \eqref{bigscales} and \eqref{quarea}, we obtain
\beq
\frac{LM}{2\pi\hbar} = N^{1/4}
\eeq
This result is unusual: we interpret it to mean that if we regard $N$ as fixed, then the cutoffs on space and momentum are not separately arbitrary but are inversely related. This clearly can be interpreted as a UV/IR mixing phenomenon. In EFT, there is no such relation, because there is no notion of finite $N$. 

Let us expand on this point. As we have seen above, 
the vacuum partition function $Z_{S^1}$ for a particle scales as the product of the volume of spacetime and the volume of momentum space, i.e., as the
covariant \emph{phase space volume}. It is natural to associate this volume with the counting of a number of states. In the classical particle picture, there is one state for each value of the position in spacetime and each value of 4-momentum. Naively, the number of states is, therefore, infinite. Both UV and IR regulators are needed to make the phase space volume finite, where it becomes the product of two factors, $V_q$ and $V_p\sim M^4$. In quantum theory, there is a bound on how many states we can put in a given finite size phase space, since $\hbar$ sets the scale for a minimal phase space volume. {\it Therefore, the finite phase space volume corresponds to a finite number of quantum states.} That is, with an IR and UV regulator in place, the phase space is compact and  the Hilbert space is finite-dimensional \cite{BateWei}. Here $N$ is the dimension of the Hilbert space.

Returning to the regulated circle amplitude \eqref{fullZtaumr}, we then find the corresponding regulated energy density from \eqref{relZrho},
\beqn\label{zellmr}
\rho(\tau)_{m.r.}=\hbar\Big[
\frac{\varepsilon N_p}{2\pi\hbar }\frac{1}{N_q}
\sum_{ k=0}^{N_{q}-1}\int_0^1 d\tilde x\, e^{-(\tilde x+k)^2\varepsilon^2\tau/2}\Big]^4.
\eeqn
We note that all states that make an appreciable contribution are such that the argument of the exponential in \eqref{zellmr} is small. The $\tau$-integration is not expected to change the scaling, and so we arrive at an upper bound on the vacuum energy density \beq
\rho_{m.r.}\lesssim 
\hbar\Big[\frac{\varepsilon N_p}{2\pi\hbar}\Big]^4
=\hbar\Big[ \frac{M}{2\pi\hbar}\Big]^4.
\eeq
This result is no surprise, as $M$ is the total size of regulated momentum space; if it is identified as a large mass scale such as $m_{P}$, then the usual conundrum pertains. 

On the other hand, given the
quantum area constraint, the bound on the energy density can be rewritten as
\beq\label{rhoboundraw}
\rho_{m.r.}\lesssim 
\hbar\frac{N}{V_q}
\eeq
where $V_q\simeq L^4$.  
As far as we are aware, this formula is new. We note that it relates the vacuum energy density times space-time volume, $\rho V_q$, to $N$. Given the form of this relation, if we interpret $N$ as a count of microscopic degrees of freedom, 
 it is natural to say that $N$ should be interpreted as an entropy, in which case the bound on the vacuum energy density resembles an `equation of state', $\rho V\sim S$. Indeed, in what follows we will equate this entropy with a macroscopic notion of gravitational entropy. This makes a direct connection between the microscopic calculation of the vacuum energy density and (quantum) gravitational physics and as we will see yields a bound on the vacuum energy density that is remarkably consistent with observations. 




We reiterate that if we were to identify $\lambda$ with the Planck length $\ell_P=\sqrt{\hbar G_N}$, and correspondingly $\varepsilon\sim m_P = \sqrt{\hbar/G_N}$, then this bound corresponds to the usual EFT result $\rho\sim m_P^4$.
However, there is nothing in this construction that forces us to take $\lambda\sim \ell_P$; $\lambda$ and $\varepsilon$ are constrained only by the quantum phase space area constraint $\lambda\varepsilon=2\pi\hbar$. In ordinary quantum mechanics similarly, the shape of a phase space cell 
is determined contextually by some experimental setup. For example, in the context of a double slit experiment the  length scale $\lambda$ is  the distance between the slits  \cite{aharonov}. 
Often in quantum gravity discussions, some notion of minimal volumes or areas in space-time is introduced. All such notions are {\it ad hoc}, while here we are introducing minimal volumes in phase space which have a natural interpretation in quantum theory.


Nevertheless, it is well known that in gravity there is a concept of entropy which scales as area rather than the volume of a spatial subregion
\cite{Bekenstein:1973ur},
\cite{hawking}. Furthermore in any theory of quantum gravity, it is commonly suggested that this should be thought of as entanglement entropy \cite{Harlow:2022qsq}. Below, we will argue that the missing ingredient in EFT is this non-extensivity of entropy. We suggest that the holographic bound on entropy can be thought of as providing a contextual scale that determines the shape of elementary phase space cells. 

To do so, we will identify $N$ with the gravitational entropy
\beq
S_{grav} = \ell_P^{-2} Area\sim (\ell/\ell_P)^2
\eeq
where $\ell$ is some characteristic length scale. 
This identification  can also be phrased in terms of
 a Bekenstein bound \cite{Bekenstein:1980jp,Bousso:2002ju,Marolf:2003sq,Casini:2008cr}, but here we will need only a simple implementation:
 we are interested in a space-time with a positive vacuum energy density, in other words, an asymptotically de Sitter space-time with its cosmological horizon \cite{hawking}. In this case the Bekenstein bound gives the  entropy $S_{grav}$ of the cosmological horizon in four space-time dimensions.

Thus we identify the length scale $L$ with the size of the whole observable Universe $\ell$,  the cosmological horizon,  $\ell \equiv L= N_q\lambda$.
Equating $S_{grav.}$ with $N$, eq. \eqref{rhoboundraw}  gives 
\beq
N \sim \ell^2/\ell_P^2
\eeq which gives us a bound on energy density given by
\beq
\rho
\lesssim \hbar \left(\frac{\ell_P}{\ell}\right)^2\ell_P^{-4} = \frac{\hbar}{\ell^2 \ell_P^2}
\eeq
or equivalently (by looking at the Einstein-Hilbert action on-shell), a value for the cosmological constant
\beq
\Lambda_{c.c.}\lesssim 
\, 8\pi G_N \rho
\sim \frac{1}{\ell^2}
\eeq
{Note that this result does not depend on the characteristic UV scale represented by $\ell_P$, but only depends on the characteristic IR scale $\ell$.
This result is also technically natural, because the value of the cosmological 
constant goes to zero as the IR size increases to infinity. The factor $(\ell_P/\ell)^2$ provides the needed
hierarchy between the observed energy density and the naive prediction of EFT.}

This result is vastly smaller than the naive effective field theory computation. What has happened? We interpret this result to imply that the holographic bound on entropy implies a corresponding bound on the vacuum energy density. Apparently, the naive effective field theory reasoning overestimates the available phase space by a large amount, assuming implicitly that entropy is extensive. This can be stated in terms of an effective momentum space size which is set by $\varepsilon N_p$. Note that whereas $N_q$ is very large, being associated with the size of the cosmological horizon, $N_p$ is unconstrained and for simplicity can be taken to be order one. Thus the effective momentum space size is set approximately by
$
M\sim  \varepsilon
$
and given the above result, we see that
\beq
M\sim \hbar \ell_P^{-1} (\ell_P/\ell)^{1/2} {
= \frac\hbar{\sqrt{\ell\ell_P}}
}.
\eeq
Interestingly, this is a macroscopic scale. It implies that the typical phase space cell is much wider in the space direction and much thinner in the momentum direction 
than one would guess by identifying them with the Planck scales. One might regard the gravitational entropy as providing a context for setting these scales. 
Indeed, if we had an actual theory of quantum geometry to which we could couple a quantum field theory, we should expect that this matching of entropies would be automatic. 


To complete the discussion, let us provide some realistic numbers.  
The Hubble scale of the visible universe is $\ell \sim 10^{27}m$, the Planck length is $\ell_P \sim 10^{-35}m$, and thus 
$
N\sim \ell^2/\ell_P^2 \sim 10^{124}.
$ 
This formula summarizes our main point which is that   the size of the universe is simply proportional to the number of degrees of freedom in the universe  and that the universe is large because it contains a lot of degrees of freedom. Indeed if the universe contains $N$ degrees of freedom the fluctuations scale as $1/\sqrt N$, and the stability of the universe condition that the fluctuations are relatively small translates to the condition that $N$ is large \cite{erwin}.
   The observed
vacuum energy density \cite{Riess:1998cb, Perlmutter:1998np} is $[10^{-3}eV]^4$, and this value is indeed
$10^{124}$ smaller than the naive EFT estimate corresponding to the Planckian vacuum energy density $[10^{19} GeV]^4$.
The above geometric mean of the Hubble scale and the Planck length is $\lambda \sim (\ell \ell_P)^{1/2} \sim 10^{-4}m$, 
which corresponds to the observed vacuum energy scale of $M \sim 10^{-3}eV$.


\bigskip

In this paper we have introduced the concept of modular regularization \eqref{fullZtaumr}.  In a quantum theory, the trace that yields the vacuum energy density will be basis-independent; any two choices of basis are unitarily related (for example, the momentum polarization and the modular polarization are related by Zak transform \cite{Freidel:2016pls}). In EFT though, we are tied to a preferred basis, having assumed a {\it fixed classical space-time}. Regulating and performing the trace in terms of a modular polarization
might then be interpreted as introducing a quantum notion of space-time geometry. The reader may find it confusing that we have identified the modular cells with ground states, as in EFT the ground state is unique. This is not inconsistent though: if the momentum polarization is thought of as a singular limit in which the modular cell is squashed, $\varepsilon\to 0$ and $\lambda\to\infty$, the information about multiple cells is simply lost in the limit. Thus by interpreting the trace as a phase space integral in a modular polarization, we introduce a regulator that retains the short-distance quantum structure of the phase space.
On the other hand, most EFT calculations (such as of correlation functions) are unaffected by such considerations, where information about the vacuum energy simply factors out. But for computations that do directly address vacua, EFT simply has no mechanism to access a ground state degeneracy. In standard treatments where we have classical space-time geometry, one would attempt to calculate an entropy by referring to a quantum field theory Hilbert space; it would be interesting to connect our notion of ground state degeneracy with a regularization of such a calculation.


In this short paper, we have chosen to present the concepts in the most straightforward terms. There however is no obstruction to implementing them in string theory or other quantum gravitational contexts  \cite{deBoer:2022zka}.
For example, we can implement the same construction in the vacuum energy computation in perturbative string theory \cite{Polchinski:1998rq, Polchinski:1985zf}. It seems possible that this can be interpreted in terms of duality-covariant constructions of string theory 
\cite{Freidel:2013zga, Freidel:2014qna,Freidel:2015pka, Freidel:2017wst, Freidel:2017nhg, Freidel:2018apz,Freidel:2016pls}.
In such a formulation, a four-dimensional space-time emerges through a process of extensification \cite{Freidel:2016pls}, with compact directions remaining of Planckian size and not affecting the vacuum energy. 

Predicting a suitably small vacuum energy is only part of the cosmological constant problem. It is well known that there is another aspect, 
the coincidence problem \cite{Weinberg:2000yb}. Having tied the vacuum energy density to the size of the cosmological horizon, this problem may be  resolved as well. It will also be interesting to consider if other hierarchy problems might be treated by similar methods; in the context of string theory, we note that there is a relation between the cosmological constant and the gauge
hierarchy problem \cite{Abel:2021tyt}.


{\bf Acknowledgments}: We thank 
P. Berglund, S. Goldman, T. H\"{u}bsch, W. Jia, M. Klinger,
D. Mattingly, P.-C. Pai and T. Takeuchi for discussions. The work of RGL was supported by the U.S. Department of Energy under contract DE-SC0015655 and DM by the U.S. Department of Energy under contract DE-SC0020262.
For JKG, this work was supported by funds provided by the National Science Center, project number  2019/33/B/ST2/00050.
Research at Perimeter Institute for Theoretical Physics is supported in part by the Government of Canada through NSERC and by the Province of Ontario through MRI.

\end{document}